\documentstyle[aps,twocolumn]{revtex}
\input epsf
\begin{document}
\draft
\title{Temperature and spatial diffusion of atoms cooled in a 3D 
lin$\perp$lin bright optical lattice}

\author{F.-R. Carminati, M. Schiavoni, L. Sanchez-Palencia, 
        F. Renzoni and G. Grynberg}

\address{Laboratoire Kastler-Brossel, D\'epartement de Physique de l'Ecole
Normale Sup\'erieure, 24 rue Lhomond, 75231, Paris Cedex 05, France}

\date{\today{ }}

\maketitle
\begin{abstract}
We present a detailed experimental study of a three-dimensional
lin$\perp$lin bright optical lattice.
Measurements of the atomic temperature and spatial diffusion coefficients 
are reported for different angles between the lattice beams, i.e. for
different lattice constants. The experimental findings are interpretated
with the help of numerical simulations. In particular we show, both 
experimentally and theoretically, that the temperature is independent of 
the lattice constant.
\end{abstract}

\section{Introduction}

The spatial modulation of the intensity or polarization of laser
beams leads to cooling and trapping of atoms
\cite{grimm,dalibard89}. In the case of a {\it periodic} spatial
modulation of the light, the trapping of atoms results in
their localization in a periodic structure, called optical lattice
\cite{jessen,robi}.

Near-resonant bright optical lattices, the ones considered in this work, are 
based on the Sisyphus cooling mechanism \cite{dalibard89,castin91}. The 
periodic modulation of the light polarization, produced by the 
interference of several laser beams, leads to a periodic modulation
of the light shifts (optical potentials) of the different Zeeman sublevels 
of the ground state of the atom. As a result of the optical pumping 
between different optical potentials, atoms are cooled and finally trapped 
at the potential minima. 

The lin$\perp$lin configuration, which corresponds to a Sisyphus cooling, and 
the $\sigma^{+}-\sigma^{-}$ one have been the first laser cooling schemes
leading to sub-Doppler temperatures \cite{dalibard89}. 
Therefore much experimental and theoretical work has been done on these 
cooling schemes, and many temperature measurements have been reported. 
However the majority of the experimental data corresponds to 1D configurations 
or to 3D optical molasses produced by six phase-uncontrolled laser beams
\cite{metcalf}.
A much smaller amount of experimental data is available for 3D optical 
lattices, as obtained by the interference of four phase-uncontrolled 
laser beams \cite{tetra}. Besides the early studies by our group
\cite{robi}, temperature measurements for a four-beam 3D lattice have been 
done at NIST \cite{nist97} and later on by the Kastberg's group 
\cite{kastberg00,kastberg01}. In these works the temperature of the atoms
in a lattice with fixed geometry was studied as a function of the atom-light
interaction parameters (laser intensity and detuning, corresponding to 
different depths of the potential well and different optical pumping rates).
No attention was payed to dependencies of the temperature on geometrical 
parameters as the lattice constants, which can be varied by changing the 
angles between the lattice beams. That is one of the issues analyzed in the 
present work.

\begin{figure}[ht]
\begin{center}
\mbox{\epsfxsize 3.in \epsfbox{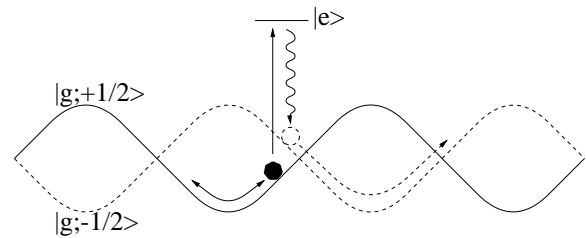}}
\end{center}
\caption{Process of spatial diffusion produced by optical pumping between
different optical potentials. The shown potentials correspond to a
$J_g=1/2 \rightarrow J_e=3/2$ transition in counterpropagating laser
fields with orthogonal linear polarizations.}
\label{hopping}
\end{figure}

Another aspect of optical lattices relatively unexplored is the diffusive
atomic dynamics. That will be the second issue analyzed in this work.
In a bright optical lattice, as obtained with a laser field red 
detuned from a $F_g=F\to F_e=F+1$ transition, atoms at the bottom of a 
potential well strongly interact with the light and therefore undergo 
fluorescence cycles. This produces complex transport phenomena which have
not yet been completely understood.  Two different mechanisms may lead to 
spatial diffusion. First, the optical pumping may transfer an atom from a 
potential well to a neighbouring one corresponding to a different optical
potential, as shown in Fig. \ref{hopping}. 
Second, atoms are heated as result of the recoil associated to the 
fluorescence cycles. 
Thus, the increase in kinetic energy following many fluorescence cycles 
may allow the atom to leave the potential well. In this case the atomic 
diffusion is not associated to optical pumping between different ground
state sublevels.

There has already been a few investigations devoted to the study of
spatial diffusion in optical molasses and bright optical lattices 
\cite{hodapp,talbot,speckle,guidoni,aspect}.
In Refs. \cite{hodapp,talbot,speckle,guidoni} the atomic transport and
spatial diffusion has been studied by a direct imaging of the expansion 
of the atomic cloud. A completely different approach has been followed 
in Ref. \cite{aspect} where both the atomic dynamics in a single
potential well and the atomic transport in the lattice have been 
studied through polarization-selective intensity correlation 
spectroscopy.

In this work we present a detailed experimental study of the temperature
and spatial diffusion of $^{85}$Rb atoms cooled in a 3D lin$\perp$lin
bright optical lattice.
Measurements for different angles between the lattice beams, i.e. for 
different lattice constants, are presented. The experimental results 
are compared with numerical simulations.

This work is organized as follows. In Section II we describe the 
experimental set-up. Section III contains the results of the 
temperature measurements obtained by recoil-induced resonances
and the corresponding numerically calculated values.
In Section IV we present our results for the measurement of the
spatial diffusion coefficients by direct imaging of the atomic cloud
and we compare them with the numerical simulations.   
Finally, in Section V we present the conclusions of our work.

\section{Experimental set-up}

The preparation of the atoms in the optical lattice is the standard 
one used in previous experiments \cite{tetra}. The rubidium atoms are
cooled and trapped in a magneto-optical trap (MOT). Then the MOT
magnetic field and laser beams are turned off and the atoms are
cooled and trapped in the optical lattice. The periodic structure
is determined by the interference of four linearly polarized laser
beams, arranged as shown in Fig. \ref{setup}: two $y$-polarized beams,
symmetrically disposed with respect to the $z$ axis, propagate in the 
$xOz$ plane with a relative angle $2\theta_x$, while two $x$-polarized 
beams, also symmetrically disposed with respect to the $z$ axis, propagate 
in the $yOz$ plane and form an angle $2\theta_y$. As discussed in Refs. 
\cite{tetra,petsas}, this configuration results in a 3D periodic potential,
 with minima located on an orthorhombic lattice and associated with pure
 circular (alternatively $\sigma^{+}$ and $\sigma^{-}$) polarization.

For the measurements presented in this work, the angles between
the lattice beams in the $xOz$ and $yOz$ planes are equal: 
$\theta_x=\theta_y\equiv\theta$. This corresponds to the lattice
constants, i.e. to the distance (along a major axis) between two
sites of equal circular polarization, $\lambda_{x,y}=\lambda/\sin\theta$,
$\lambda_z=\lambda/(2\cos\theta )$, with $\lambda$ the laser field
wavelength. For this configuration with equal angles and equal intensity
laser beams the radiation pressure is balanced.

For the MOT and the lattice we used the $F_g=3\to F_e=4$ D$_2$ line transition 
of $^{85}$Rb. A repumping beam resonant to the $F_g=2\to F_e=3$ transition
is also added.

\section{Temperature measurements}

\begin{figure}[ht]
\begin{center}
\mbox{\epsfxsize 3.in \epsfbox{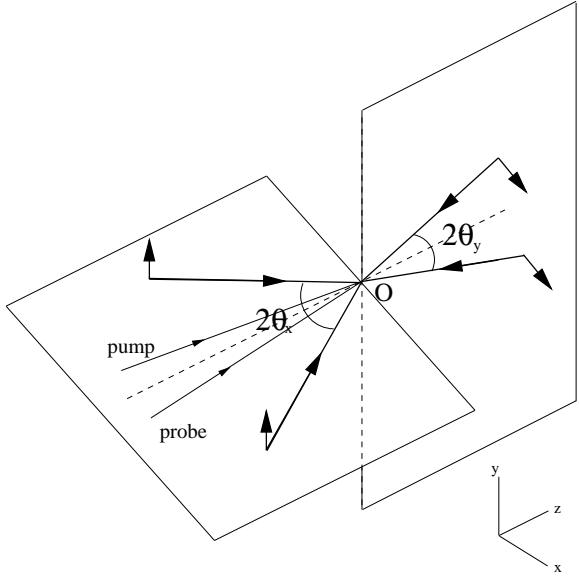}}
\end{center}
\caption{Sketch of the experimental setup. We indicate by
$2\theta_x$ ($2\theta_y$) the angle between the lattice beams
in the $xOz$ ($yOz$) plane.
For the measurements presented in this work, the angles between
the lattice beams in the $xOz$ and $yOz$ planes are equal:
$\theta_x=\theta_y\equiv\theta$. Two additional laser fields (the pump and
probe beams) are used for the temperature measurements.}
\label{setup}
\end{figure}

We measured the temperature of the atomic sample after a $20$ ms cooling
phase in the optical lattice by using the method of recoil-induced resonances 
\cite{courtois94,meacher,courtois96}. Once the lattice beams are turned off,
two additional laser beams (the pump and the probe beams, see Fig. \ref{setup})
are introduced for the temperature measurements.
They cross the atomic sample in the $xOz$ plane, and they are symmetrically 
disposed with respect to the $z$ axis forming an angle of $23^0$. The probe
transmission is monitored as a function of the detuning between the
pump and probe fields. For equal pump and probe polarizations, the states
$|\alpha, p\rangle$ and $|\alpha, p'\rangle$ with the same internal
quantum number $\alpha$ but different atomic momentum are coupled by
Raman transitions. The difference in population of these states determines
the absorption or gain of the probe, and the probe transmission spectrum
results to be proportional to the derivative of the atomic momentum
distribution. From the width of the resonance in the probe
transmission spectrum it is then straighforward to derive the atomic temperature
\cite{meacher}. The geometry of our pump and probe fields corresponds to 
measurement of the temperature in the $x$ direction \cite{remark}.

\begin{figure}[ht]
\begin{center}
\mbox{\epsfxsize 3.5in \epsfbox{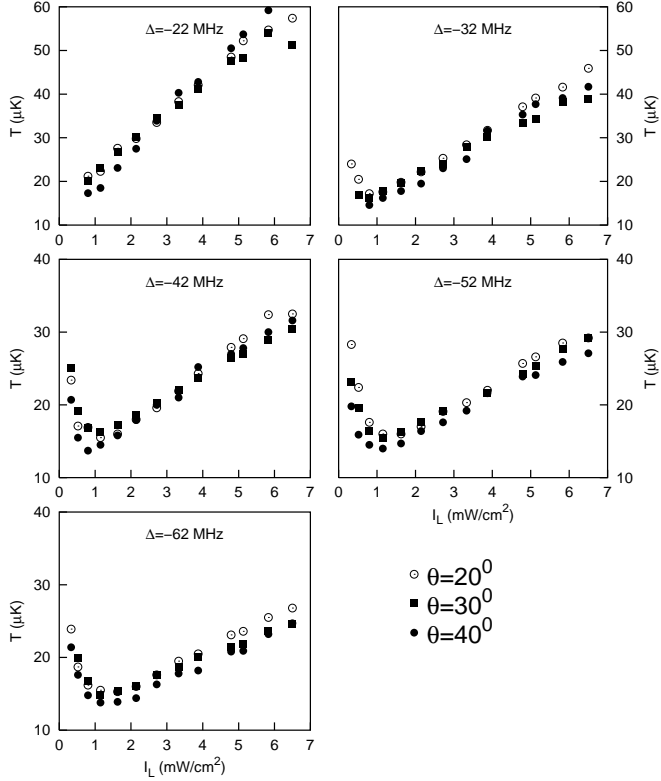}}
\end{center}
\caption{Atomic temperature in the $x$ direction as a function of the
intensity per lattice beam $I_L$ at different values of the lattice
detuning $\Delta$ and for different choices of the lattice angle $\theta$.}
\label{fig_temp}
\end{figure}

Results of our measurements for the temperature as a function of the 
lattice beam intensity at different values of the lattice detuning and for 
different choices of the lattice angle $\theta$ are shown in Fig. 
\ref{fig_temp}.

We observe a linear dependence of the temperature with the laser intensity,
with a {\it d\'ecrochage} at small intensities, in agreement with theoretical
models and previous measurements \cite{kastberg00,kastberg01}.

As an original result, on the basis of the measurement reported in Fig.
\ref{fig_temp}, we find that the temperature is independent of the 
angle between the lattice beams, i.e. of the lattice constant.
However, this behaviour, independent of the lattice constant, is specific of
the temperature and we should not conclude that the atomic dynamics does
not depend on the lattice constants.
In fact, as we will show in the following, other quantities, such as the 
spatial diffusion coefficients, show a strong dependence on the angle 
between the lattice beams. 

\begin{figure}[ht]
\begin{center}
\mbox{\epsfxsize 3.in \epsfbox{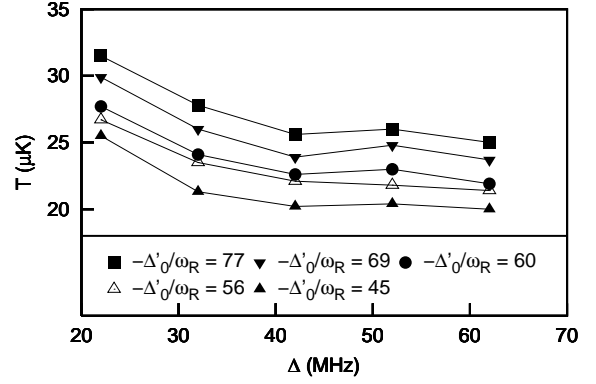}}
\end{center}
\caption{Atomic temperature in the $x$ direction as a function of the lattice
detuning $\Delta$, for different values of the light shift per beam
$\Delta_0^{'}$, {\it i.e.} at different depths of the potential well
($\omega_R$ is the atomic recoil frequency). The data refer to a lattice
angle $\theta=30^0$. The lines are guides for the eyes.}
\label{fig_temp2}
\end{figure}

We also studied the temperature of the atomic sample as a function of the
lattice detuning $\Delta$, for different depths of the potential well
$U_0$, i.e. for different values of the light shift per beam $\Delta_0^{'}$
($U_0$ is proportional to $\Delta_0^{'}$).
Results of our measurements are shown in Fig. \ref{fig_temp2}. We conclude 
that the temperature depends on the depth of the potential well only, and 
not on the lattice detuning, except for an increase at small detunings.
We have also performed a theoretical study with the help of semi-classical 
Monte Carlo simulations \cite{lsp}. 
Taking advantage of the symmetry between the $x$
and $y$ directions (see Fig. \ref{setup}), we restricted the atomic dynamics
in the $xOz$ plane. The numerical calculations in this restricted configuration
give the same dependencies of the dynamical quantities on the
lattice parameters as full 3D calculations, so they are useful to interpret
the experimental results.
However, a direct quantitative comparison is not necessarily meaningful
because the restriction of the dynamics to two dimensions may introduce
scaling factors (see \cite{lsp} for more details).

We considered $1/2\to 3/2$ atoms, as customary in numerical analyses of
Sisyphus cooling. Because the external degrees 
of freedom of the atoms are treated as classical variables, it is 
straighforward to calculate the variance of the atomic momentum 
distribution, which is proportional to the steady-state temperature in a 
thermalized cloud
\begin{equation}
k_B T_i=\frac{\langle p_i^2\rangle}{M}~(i=x,z)~,
\end{equation}
with $p_i$ the atomic momentum in the $i-$ direction, $M$ the atomic mass
and $k_B$ the Boltzmann constant.
We found results in good agreement with the experimental ones. In particular,
the independence of the temperature on the lattice angle is well confirmed 
by the numerical simulations. The physical picture of Sisyphus cooling 
\cite{dalibard89} allows an immediate interpretation of this result: an atom
looses kinetic energy until it gets smaller than the depth of the potential
well, independently of the lattice constant.
More precisely in the regime considered here (the {\it jumping} regime
\cite{dalibard89}), the atomic dynamics is well described by a brownian 
motion: an atom undergoes a friction force $F_i=-\alpha_i v_i$ and a
fluctuating force, essentially produced by the fluctuations of the dipole 
force, which is described by the momentum diffusion coefficient $D_{p_i}$
in the $i-$ direction.
As shown in Ref. \cite{lsp}, $\alpha_i$ and $D_{p_i}$ are proportional to
$1/\lambda_i^2$ in a domain where the spontaneous emission can be neglected. 
Hence, the temperature, given by the Einstein relation 
$k_B T_i =  D_{p_i}/\alpha_i$, is independent of the lattice constants.

The increase of the temperature at small lattice detuning is also present 
in our numerical results.
As the laser frequency gets close to the atomic resonance the momentum
diffusion corresponding to multiple atomic recoils increases while the 
friction coefficient associated to the Sisyphus cooling decreases 
\cite{dalibard89}. As a result, at small lattice detuning the equilibrium
temperature increases approaching the atomic resonance.  We verified this 
point by arbitrarily removing the corresponding term in the simulation. Then,
no increase of the temperature at small detuning appeared.

\section{Measurement of the spatial diffusion coefficients}

\begin{figure}[ht]
\setlength{\unitlength}{1in}
\vspace{0.2cm}
\begin{picture}(4,1.6)
\put(-0.1,0.9){\epsfxsize 1.1 in \epsfbox{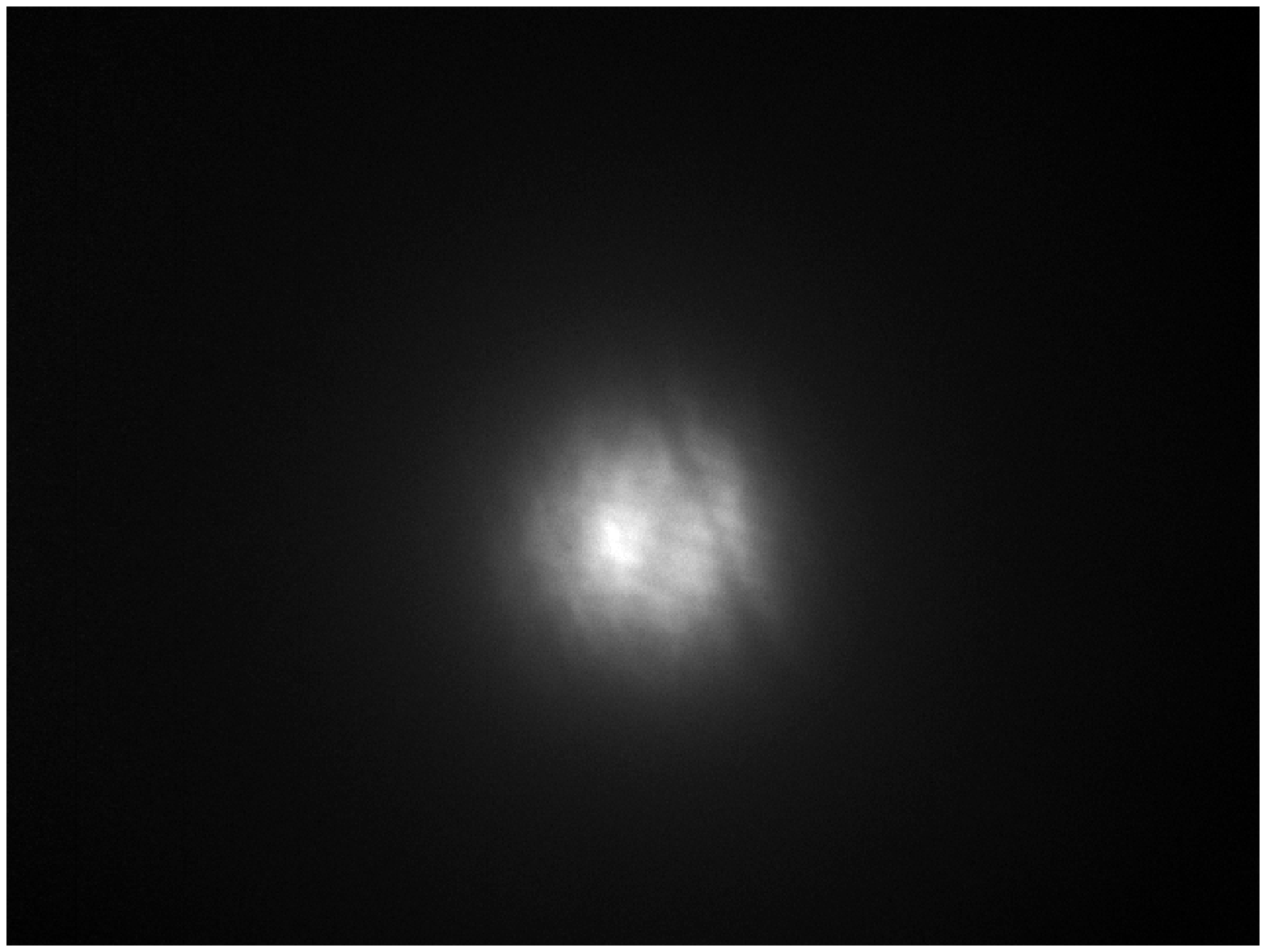}}
\put(-0.1,0.0){\epsfxsize 1.1 in \epsfbox{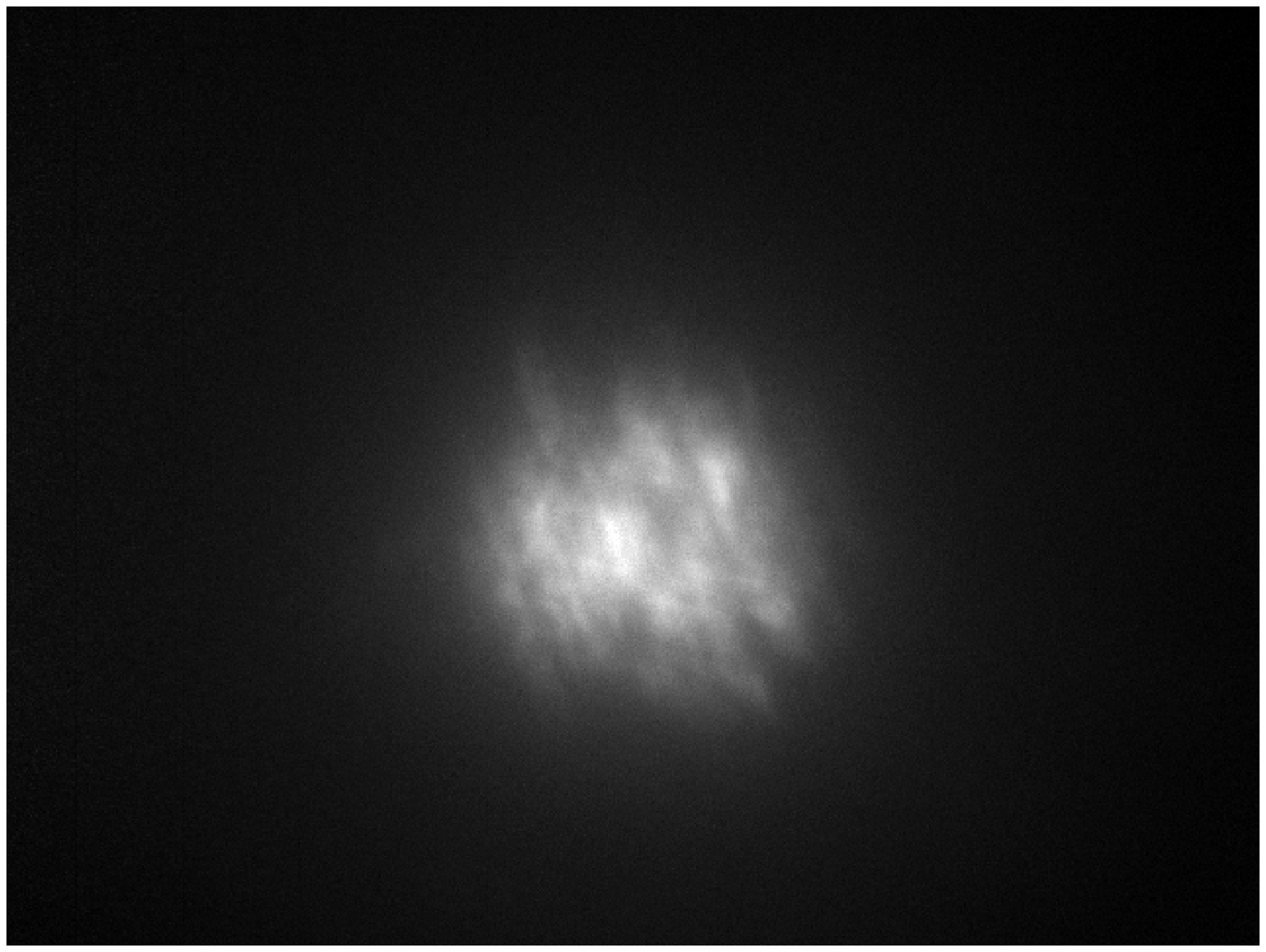}}
\put(1.2,0.1){\epsfxsize 2. in \epsfbox{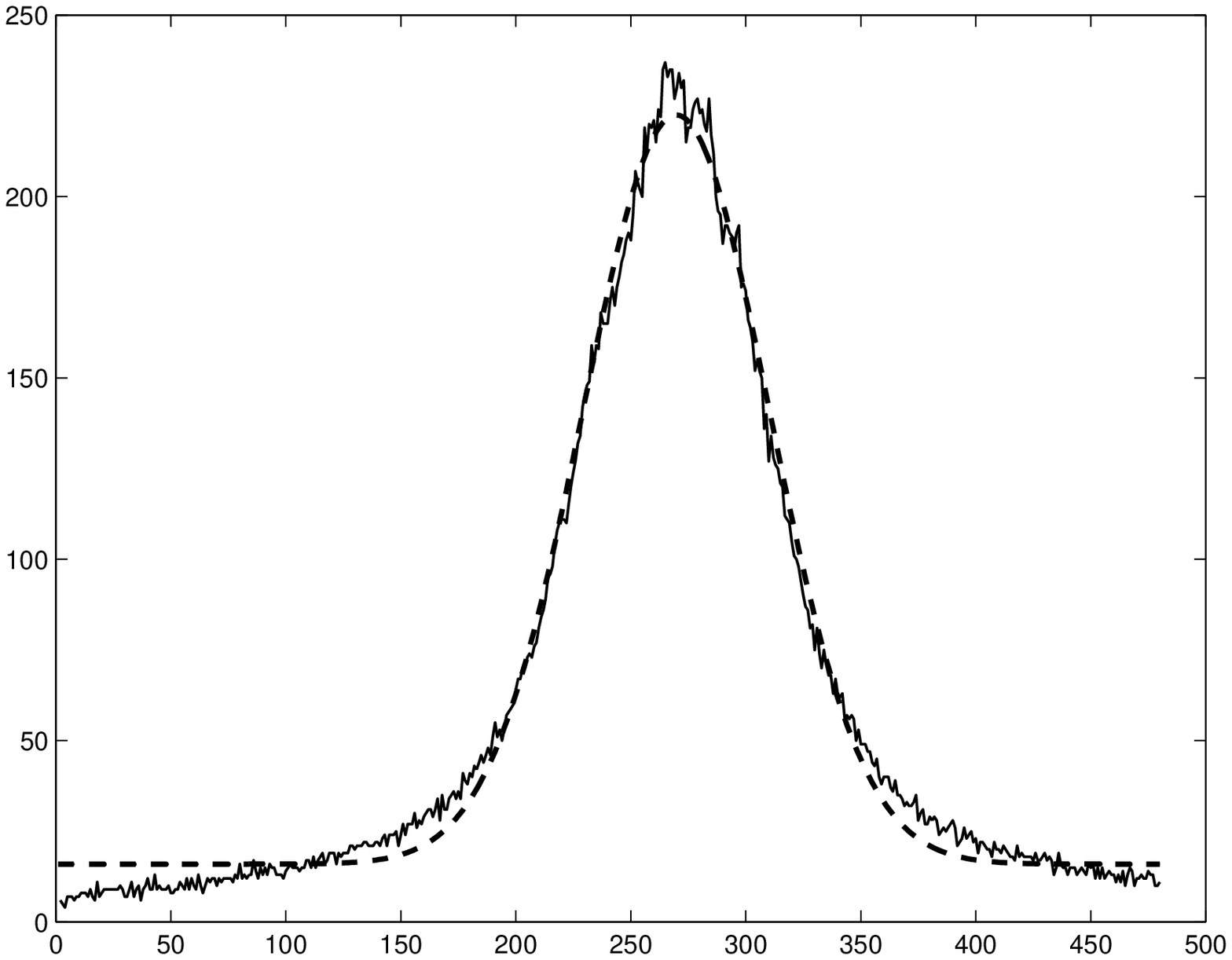}}
\end{picture}
\vspace{0.2cm}
\caption{
Left: Images of the atomic cloud after 50 $ms$ (top) and 400 $ms$
(bottom) from the loading of the optical lattice. Right: A typical
profile of the atomic cloud (continuous line). The dashed line
represents the best fit with a gaussian function.}
\label{image}
\end{figure}

We now turn to the study of spatial diffusion. We observed the 
atomic cloud expansion by using a Charge Coupled Device (CCD) camera.
We took images of the expanding cloud at different instants after
the atoms have been loaded into the optical lattice. The exposure lasts 
1 ms and we took an image every 7 ms.  Since we wanted to explore regions
of large detuning and low intensity for the lattice beams, we increased 
to maximum the lattice beams intensity while imaging the atomic cloud
to increase the brightness of the image. Of course, the atomic dynamics
was affected during the imaging and we took only an image for every
realization of the diffusion. Since the $x$ and $y$ directions are
equivalent in our lattice, we choose to take images in a plane containing
the $z$ axis and forming an angle of $45^o$ with the $x$ and $y$ axis
($\xi$ axis). Fig. \ref{image} shows typical images of the expanding cloud.

\begin{figure}[ht]
\begin{center}
\mbox{\epsfxsize 3.in \epsfbox{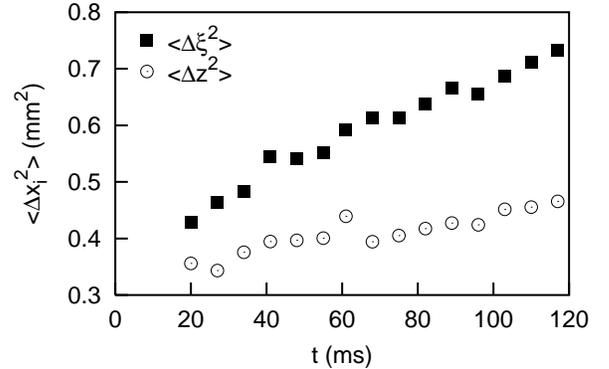}}
\end{center}
\caption{Atomic mean square displacement as a function of time.
The data correspond to a laser detuning of $-62.35$ MHz and an
intensity per lattice beam of 3.2 mW/cm$^2$.}
\label{Diff3}
\end{figure}

We verified that the profile of the atomic cloud is well approximated
by a gaussian function (Fig. \ref{image}, right). From the images of
the atomic cloud we derived the atomic mean square displacements
$\langle \Delta z^2\rangle$ and $\langle \Delta \xi^2\rangle$. As
evidenced by the plots of Fig. \ref{Diff3}, the cloud expansion 
corresponds to a normal diffusion, i.e. the mean square atomic 
displacement increases linearly with time.

\begin{figure}[ht]
\begin{center}
\mbox{\epsfxsize 3.5in \epsfbox{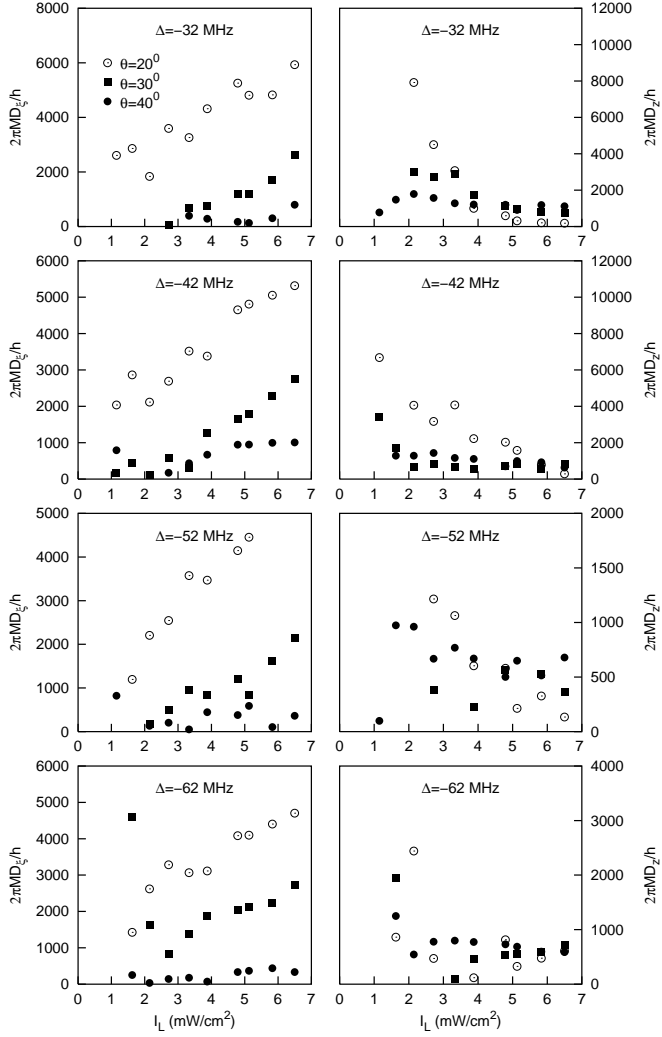}}
\end{center}
\caption{Experimental results for the spatial diffusion coefficients in the
$\xi$ and $z$-directions as a function of the laser intensity per lattice
beam for different values of the lattice detuning and different lattice
angles $\theta$.}
\label{Diff}
\end{figure}

From data as those of Fig. \ref{Diff3} we have been able to derive
the spatial diffusion coefficients $D_z$ and $D_{\xi}$ in the $z$- and 
$\xi$-directions by fitting the experimental data with
\begin{equation}
\langle \Delta x_i^2 \rangle = 2D_i t~~~~(i=z,\xi)~.
\label{coeff}
\end{equation}
We determined the spatial diffusion coefficients as a function of the 
lattice beam intensity for different values of the laser detuning and 
lattice constant. Our results are shown in Fig. \ref{Diff}.

It appears that $D_{\xi}$ is an increasing function of the lattice
intensity. On the other hand, the experimental data do not show a clear
general dependence of $D_z$ on the laser intensity.

As an important fact, we observe that $D_\xi$ is a decreasing function 
of the angle $\theta$ between the lattice beams. This supports the picture
of the atomic spatial diffusion as produced by optical pumping (see Fig. 
\ref{hopping}). Indeed if we assume that the diffusion is produced by the
optical pumping between neighbouring wells, the diffusion coefficients
(Eq. \ref{coeff}) will be simply proportional to the pumping rate and to 
the square of the lattice constant. As the lattice constant 
$\lambda_{x,y} = \lambda/\sin\theta$ decreases for increasing $\theta$, 
the spatial diffusion coefficients decreases as a result.

It is difficult to test whether a similar argument applies also to the 
diffusion in the $z$ direction, as the range of 
$\lambda_z=\lambda/(2\cos\theta )$ explored experimentally has been
quite small because of practical difficulties with lattice beams with
very large or very small angles. However we note that at large lattice 
intensity, the spatial diffusion coefficient in the $z-$ direction 
increases for increasing $\lambda_z$, i.e. for increasing $\theta$, as 
expected. This behaviour is not observed at small lattice intensity. 

\begin{figure}[ht]
\begin{center}
\mbox{\epsfxsize 3.5in \epsfbox{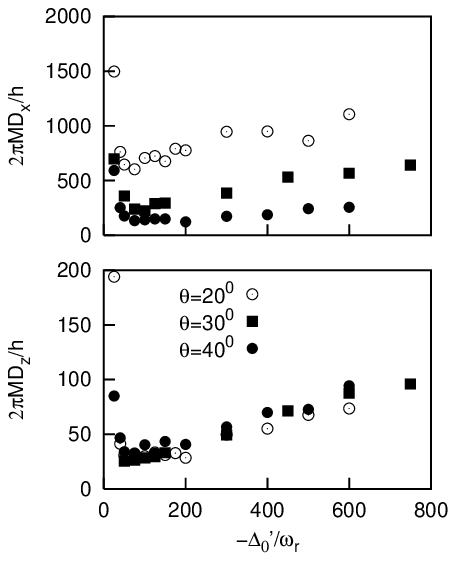}}
\end{center}
\caption{
Numerical results for the spatial diffusion coefficient in the $x$ and
$z$-directions as a function of the light shift per beam at fixed lattice
beam detuning ($\Delta = - 3\Gamma$, with $\Gamma/2\pi = 5.9$ MHz for
$^{85}$Rb), i.e. as a function of the lattice beams intensity.
The different data sets correspond to different lattice angles $\theta$.}
\label{diffnum}
\end{figure}

In the numerical simulations, we monitored the variance of the atomic 
position distribution as a function of time. We verified that the 
spatial diffusion is normal, and determined the diffusion coefficients
from Eq. \ref{coeff}. The results of our calculations are shown in 
Fig. \ref{diffnum}.

We note a linear dependence of the diffusion coefficients on the lattice
beams intensity, and a {\it d\'ecrochage} at small intensity. For the values
of lattice intensity at which this {\it d\'ecrochage} is observed, the 
temperature is still in the linear regime ($T$ proportional to $\Delta_0^{'}$).
This corresponds to a transition to anomalous diffusion \cite{zoller}, 
where only few atoms, flying over several potential wells, contribute
to the diffusion process.

In the linear regime, $D_i$ is proportional to $\lambda_i^2$ ($i=\xi,z$), 
in agreement with the simple picture presented above and with the 
experimental results along the $\xi$ axis. On the contrary, in the regime of 
{\it d\'ecrochage}, the atoms do not move from a well to a neighbouring one
so no variation of $D_i$ proportional to $\lambda_i^2$ is expected. Both
the experimental and theoretical results confirm that  $D_i$ is not 
proportional to $\lambda_i^2$ in this regime.

\section{Conclusions}

In conclusion, we studied, experimentally and theoretically, the temperature
and the spatial diffusion of rubidium atoms cooled in a 3D lin$\perp$lin 
optical lattice.
The atomic temperature and the spatial diffusion coefficients are studied
for different angles between the lattice beams, i.e. for different lattice
constants. The experimental findings are interpretated with the help of 
numerical simulations. 

We show, both experimentally and theoretically, that the temperature is 
independent of the lattice constants. This is consistent with a simple
physical picture: an atom looses kinetic energy until it gets smaller 
than the depth of the potential well, independently of the lattice constants.

The experimental results for the spatial diffusion coefficients show 
a linear dependence on the lattice beams intensity, and a {\it d\'ecrochage}
at small intensity. Both  {\it d\'ecrochage} and linear regime are confirmed
by the numerical simulations. However in the experiment the {\it d\'ecrochage}
in the different directions does not appear at the same values of lattice 
parameters, in contrast with the theoretical predictions.
We have not been able to determine with certainty the origin of this
discrepancy. The choice of the $1/2\to 3/2$ transition for the numerical 
simulations, different from the one of the experiment, could be a source
of disagreement between the theoretical results and the experimental
findings. This point requires further analysis.

In the linear regime, the numerical simulations show that the spatial diffusion
coefficient is proportional to the square of the lattice constant,
in agreement with the picture of atomic diffusion produced by optical 
pumping.  Experimental findings for the diffusion coefficients along the 
$\xi$ axis and also along the $z$ axis for much higher intensities that
that leading to {\it d\'ecrochage} confirm the theoretical predictions.

We are grateful to Franck Lalo\"e for his continuous interest in our work.
We also wish to thank C.~Mennerat-Robilliard and S.~Guibal for fruitful 
discussions.
This work was supported by the European Commission (TMR network "Quantum
Structures", contract FMRX-CT96-0077).
Laboratoire Kastler Brossel is an "unit\'e mixte de recherche de l'Ecole
Normale Sup\'erieure et de l'Universit\'e Pierre et Marie Curie associ\'ee au
Centre National de la Recherche Scientifique (CNRS)".

\begin{thebibliography}{99}
\bibitem{remark}
We note that the temperature in $z$ direction can be
different from the ones in the (equivalent) $x$ and $y$
directions. In fact, as it will be discussed in detail in
the following, the temperature depends on the depth of the
potential well. Therefore in the case of anisotropy of the
optical potentials, the temperatures in the different
directions are different.
\end{thebibliography}
\end{document}